\documentclass{ws-p8-50x6-00}

\usepackage{psfig}

\def\mpc {h^{-1} {\rm Mpc}}
\def\etal{{\it et al}}
\def\eg{{\it e.g. \,}}

\newcommand{\beq}{\begin{equation}}
\newcommand{\eeq}{\end{equation}}
\def\De{{\cal D}}
\font\twelveBF=cmmib10 scaled 1000
\newcommand{\rd}{\hbox{\twelveBF x}}

\def\Journal#1#2#3#4{{#1} {\bf #2}, #3 (#4)}

\def\NAT{\em Nature}
\def\APJ{\em ApJ}
\def\MNRAS{\em MNRAS}

\def\be{\begin{equation}}
\def\ee{\end{equation}}
\def\bea{\begin{eqnarray}}
 \def\eea{\end{eqnarray}}
\begin{document}

\title{CLUSTERING IN DEEP (SUBMILLIMETRE) SURVEYS}

\author{ENRIQUE GAZTA\~NAGA \& DAVID H. HUGHES}

\address{Instituto Nacional de Astrof\'{\i}sica, \'Optica y 
Electr\'onica (INAOE), \\ Luis Enrique Erro 1, Tonantzintla, Cholula, 
78840 Puebla, Mexico}


\maketitle
\abstracts{Hughes \& Gazta\~naga (2001, see article in these 
  proceedings) have presented realistic simulations to address  key issues
  confronting existing and forthcoming submm surveys.  An important aspect
  illustrated by the simulations is the effect induced on the counts by the
  sampling variance of the large-scale galaxy clustering. We find factors
  of up to $\sim 2-4$ variation (from the mean) in the extracted counts from
  deep surveys identical in area ($\sim 6$\,arcmin$^{2}$) to the SCUBA surveys
  of the Hubble Deep Fields (HDF) \cite{h98}. Here we present a recipe to
  model the expected degree of clustering as a function of sample area and 
  redshift.}

\section{A model for the angular clustering}

\begin{figure}
\centerline{\psfig{figure=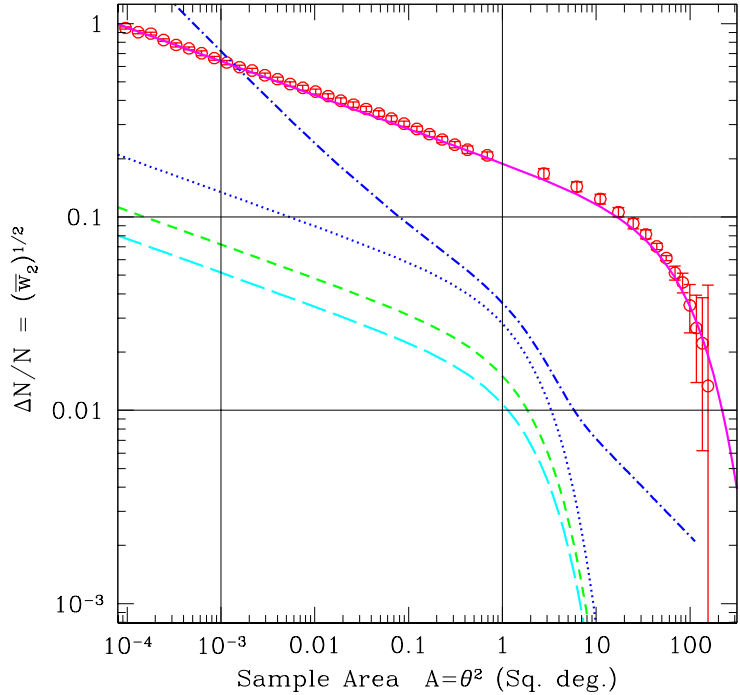,width=10 cm,angle=0}}
%\vspace{5.5cm} 
%\special{psfile=gazta_fig1.ps hscale=35 vscale=35 angle=0 hoffset=50 voffset=-40}
\caption{Mean square root deviation in number counts as a function
  of the sampled area in the sky. 
Circles with error-bars correspond to counts in square cells of
  different areas measured in the APM survey.
 The solid curve shows the model in
Eq.[\ref{eq:fitw2}]. 
The predicted fluctuations, scaled to a depth of $\De \simeq 2500 \mpc$
($z \sim 3$), are shown for different clustering evolution models:
fixed in co-moving coordinates (dotted line), stable clustering
(dashed lines) and linear theory (long-dashed lines).
The dot-dashed line includes the shot-noise contribution for
a surface density of 2000 sources/sq. deg.}
\label{w2sample}
\end{figure}

Fig.\ref{w2sample} shows fluctuations in the galaxy counts as measured in
the APM Galaxy Survey \cite{m90}. Symbols with errorbars show the
square root of the variance $\bar w_2(\theta)$, \eg $\Delta N/ N \equiv
\sqrt{\bar w_2}$, measured in squared cells of area $A \equiv \theta^2$. 
%(from Fig.1 in Gazta\~naga 1994). 
The data can be described by a power law: $$ {\Delta N / N} \equiv
\sqrt{\bar w_2} \simeq ~(A/A_0)^\beta $$ with $A_0 \simeq 7.6 \times 10^{-5}$
sq.deg. and $\beta \simeq -0.175$, for $A < 2.7$ sq. deg. For larger areas
there is an exponential cut-off with a characteristic scale of $A_c \sim 110$
sq. deg. The complete model is represented by: \beq {\Delta N \over{N}}
\simeq ~(A/A_0)^\beta ~\exp[-A/A_c],
\label{eq:fitw2}
\eeq 
and is shown as a solid curve in Fig.\ref{w2sample}.

\section{Shot-noise}

The effects of shot-noise on the counts, \eg due to the small number of sources
in the sample, is easy to quantify.  If the mean number density of sources at a
given flux is ${\cal N}$, then the number of sources in a sample of area $A$
is: $N = A {\cal N}$.  The total variance on such area is: 
\beq 
\bar w_2^{total} = \bar w_2 + {1\over{N}}, 
\label{eq:shotnoise1}
\eeq
where $\bar w_2$ is the intrinsic variance (in the case of high density) \cite{g94}.
Thus, the number counts variations due to intrinsic clustering 
and shot-noise is: 
\beq {\Delta N\over{
    N}} \simeq \left[~\bar w_2[~\theta; z; \De] + {1\over{A {\cal N}}}
\right]^{1/2}
\label{eq:shotnoise}
\eeq
where $\bar w_2[~\theta; z; \De]$ is the angular variance of depth $\De$ at redshift $z$. We next need to quantify the effects
of projection and clustering evolution.

\section{Projection effects}

Let us first assume that the 2-pt function $\xi_2(r_{12})$
 does not evolve in co-moving coordinates: $\xi_2(r_{12})=\xi_2(x_{12})$,
where $r_{12}=x_{12}/(1+z) $.
We then have that:

\beq
{\bar w_2 } (\theta)= \int_V d\rd_1 d\rd_2~
\psi(x_1)~\psi(x_2)~ \xi_2(\rd_{12}),
\label{wbarj}
\eeq
where $d\rd_i$ is the co-moving volume, 
$\psi(x)$ is the normalized
probability that a galaxy at a coordinate $x$ is included
in the catalogue and $V$ is a cone (or pyramid) of radius (or side) 
$\theta$ and infinite depth (with solid angle equal to the sample area $A$).  
In the small angle approximation, \eg  when the transverse distances 
are much  smaller  than  the radial ones, we can
relate the angular clustering 
of samples projected at two different characteristic co-moving 
depths $\De_2$ and $\De_1$ by:
\beq
{\bar w_2}\left[~\theta ~;~ \De_1~\right] = {\De_2\over{\De_1}}~{\bar w_2}\left[~\theta \De_2/\De_1 ~;~ \De_2~\right]. 
\label{eq:pro}
\eeq
We take $\De_2 \simeq 400 \mpc$ as
the mean APM depth and  $\De_1 \simeq 2500 \mpc$ as the mean co-moving 
depth of a  typical submm sample
(the actual number depends on both the mean redshift and the
cosmological parameters, as given by the luminosity distance relation).
Thus the curve that fits the APM measurements (solid curve) in 
Fig.\ref{w2sample} must be moved left by a factor $\De_2/\De_1$
to account for the fact that, at larger radial distances, the same physical
length subtends a smaller angle. The APM curve must also be moved down by a 
factor $\De_2/\De_1$ because more galaxies are seen in projection at
greater radial distances. The resulting prediction in this case,
scaled from Eq.[\ref{eq:fitw2}] to $\De_1 \simeq 2500 \mpc$ with  
Eq.[\ref{eq:pro}], is shown as the dotted line in Fig.\ref{w2sample}. 
The dotted-dashed line includes the shot-noise contribution
in Eq.[\ref{eq:shotnoise}] for a submm 850$\mu$m  flux limit of 3 mJy, with 
surface density $ {\cal N}= 2000$/sq. deg., which are characteristic
of the HDF counts. Note how in this case the variance of the counts are
dominated by shot-noise at almost all scales. These predictions
agree well with the source-count variations found in our submm simulations of
the HDF (see Hughes \& Gazta\~naga in these proceedings), which
also have clustering fixed in co-moving coordinates and the same
mean depth.

\section{The redshift evolution of clustering}
\label{sec:clusevol}

We next model the redshift evolution of $\xi_2$ in proper
coordinates $r_{12}=x_{12}/(1+z) $ as:
\beq
\xi(r_{12} ; z) ~\simeq~ (1+z)^{-(3+\epsilon)}~~ \xi(r_{12}; 0)
\label{evxi2}
\eeq 
For stable clustering (pattern fixed at proper separations) we have $\epsilon
\simeq 0$ at small scales \cite{g95}.  For a power-law correlation with slope
$\gamma \simeq 1.7$, linear theory gives $\epsilon = \gamma-1 \simeq 0.7$ 
and non-linear growth gives $\epsilon \simeq 1$.  
If the clustering is fixed in co-moving
coordinates, then $\epsilon \simeq \gamma-3 \simeq -1.3$ which produces even
less evolution than in the stable clustering regime.  This fixed clustering
model describes galaxies which are identified with high density-peaks: peaks
move less than particles, which results in less evolution.  Weak evolution is
consistent with the strong clustering observed in Lyman-break galaxies at $z
\simeq 3$, which is comparable to the clustering of present-day galaxies \cite{giav98}.  
The above model of cluster evolution projects as:
\beq {\bar
  w_2}\left[~\theta ; z_1 ; \De~\right] \simeq
\left({1+z_2\over{1+z_1}}\right)^{3+\epsilon-\gamma}~ {\bar w_2}\left[~\theta
  ; z_2 ; \De ~\right]
\label{eq:wz}
\eeq
where $\gamma$ varies  from $\simeq 1.6$ over the smaller scales
to $\gamma \simeq 2$ near the exponential break.
Here we use $z_2 \simeq 0.15$ for the APM depth
 and $z_1 \simeq 2$ for the depth in sub-mm
samples. The resulting predictions,
scaled from Eq.[\ref{eq:fitw2}] to $\De_1 \simeq 2500 \mpc$,  are shown 
as the dotted ($\epsilon = \gamma-3$), short-dashed line
 ($\epsilon = 0$) and  long-dashed lines ($\epsilon = \gamma-1$)
in Fig.\ref{w2sample}.  

\section{Discussion}

In summary, our recipe for the angular variance of number count fluctuations $
({\Delta N\over{ N}})^2 \equiv \bar w_2$ in a galaxy sample of area
$A=\theta^2$, mean co-moving depth $\De$ 
and mean redshift $z$, is given by:
\beq \bar w_2[~\theta; z; \De] \simeq ~
{400 \mpc \over{\De}}~\left({1.15\over{1+z}}\right)^{3+\epsilon-\gamma}(\theta/\theta_0)^{2\beta}
~e^{-(\theta/\theta_c)^2} ~+~ {1\over{\theta^2 {\cal N}}} 
\eeq 
where $\beta \simeq -0.175$, $\theta_c \sim 10.5$ deg., $\theta_0 \simeq 8.7
\times 10^{-3}$ deg., $3+\epsilon-\gamma$ is between 0 and 2, depending on
clustering evolution, and ${\cal N}$ is the mean number density of sources at
the given flux that produces shot-noise fluctuations (\eg
Eq.[\ref{eq:shotnoise}]).  Several cases for the above model are shown in
Fig.\ref{w2sample}.  Assuming the shot-noise is negligible, \eg ${\cal N}
\rightarrow \infty$, to reach a 1\% level of fluctuation in $N$ (lower
horizontal line in the figure) we require a sample of about $A \simeq 6$ sq.
.deg. for the model with co-moving evolution and about $A \simeq 1$ sq. deg.
for a model with strong clustering evolution. In the submm survey of the 
HDF \cite{h98}, with $A=0.001$ sq. deg.  (the left vertical line in the Fig.\,1),
we expect ${\Delta N / N} \simeq 0.2$ in the case of weak clustering
evolution and $\simeq 0.05$ for the strong evolution case.  Thus, in
principle, comparing number counts in a few more submm surveys, similar in
area to the HDF, would provide a clear discrimination between clustering
evolution models.

Nevertheless in small submm surveys the number counts for bright sources 
will be low,  \eg $N= \theta^2 {\cal N} \simeq 3$ for $S_{850\mu\rm m}
\simeq 3$\,mJy in the HDF survey \cite{h98}, 
and hence the shot-noise correction 
dominates and masks any evolution in the clustering.
This situation is shown as a dot-dashed line in Fig.\ref{w2sample} for the
submm HDF survey.  In this case, we could still
subtract the shot-noise contribution using Eq.[\ref{eq:shotnoise}] although 
this will introduce large uncertainties. 
Thus, in the future, it is necessary to extend
the surveys to larger areas and/or to lower flux densities in order to have a
discriminating measurement of clustering. The Gran Telescopio
Milim\'etrico/Large Millimeter Telescope \cite{lmt} will be the optimal
facility to provide this new generation of deep and wide mm surveys.

%\section*{References}


\begin{thebibliography}{99}

\bibitem{g94} Gazta\~{n}aga, E., \Journal{\MNRAS}{268}{913}{1994}.
\bibitem{g95} Gazta\~{n}aga, E., \Journal{\APJ}{454}{561}{1995}.
\bibitem{giav98} Giavalisco, M. \etal, 1998 \Journal{\APJ}{503}{543}{1998}.
\bibitem{h98} Hughes, D.H. \etal, \Journal{\NAT}{394}{241}{1998}.
\bibitem{m90} Maddox, S. J. \etal,
%Efstathiou, G., Sutherland, W. J., Loveday, J.
\Journal{\MNRAS}{242}{43P}{1990}.
\bibitem{lmt} http://www.lmtgtm.org/

\end{thebibliography}
\end{document}